\documentclass[prl,twocolumn,showpacs,amsmath,superscriptaddress,amssymb]{revtex4}

\usepackage{graphicx}
\usepackage{dcolumn}
\usepackage{bm}

\begin{document}


\title{Thermoelectricity by Perfectly 
Conducting Channels 
in Quantum Spin 
Hall Systems}

\author{Ryuji Takahashi}
 \affiliation{Department of Physics, Tokyo Institute of Technology,
Ookayama, Meguro-ku, Tokyo 152-8551, Japan}

\author{Shuichi Murakami}
\affiliation{Department of Physics, Tokyo Institute of Technology, 
Ookayama, Meguro-ku, Tokyo 152-8551, Japan} 
\affiliation{PRESTO, Japan Science and Technology Agency (JST),
Kawaguchi, Saitama 332-0012, Japan} 

\date{\today}

\begin{abstract}
Thermoelectric transport of two-dimensional 
quantum spin Hall systems are theoretically studied in narrow 
ribbon geometry. 
We find that at high temperature electrons in the bulk states dominate. However, by lowering temperature, the ``perfectly conducting" edge channels becomes dominant, and a bulk-to-edge crossover occurs.
Correspondingly, by lowering temperature, 
the figure of merit first decreases and then will increase again due to 
edge-state-dominated thermoelectric transport. 
\end{abstract}

\maketitle
Thermoelectric conversion of heat into energy is one of the challenging topics in material science.
The efficiency of thermoelectric energy converters depends on the transport coefficients of the constituent materials
through the figure of merit. The figure of merit $ZT$ is defined by 
$ZT= \frac{\sigma S^2T}{\kappa}$, \cite{Goldsmid}
where $T$ is the temperature, $\sigma$ is the electrical conductivity, $S$ is the Seebeck coefficient, and $\kappa$ is the thermal conductivity from electrons and phonons. 
Maximum efficiency of a thermoelectric conversion cycle depends on $ZT$, and the highest record of $ZT$ is on the order of unity. It is an important but challenging issue to search for thermoelectric systems with larger $ZT$. 
There have been several proposals to overcome this conflict and to optimize the thermoelectric
efficiency. 
One of the proposals is the  
phonon glass and electron crystal\cite{Slack}(PGEC). Because the 
phonon carries heat but not charge, phonon conduction reduces thermoelectric 
efficiency. 
Hence to achieve a high $ZT$,  
the system should be a bad conductor for phonons but a good conductor for electrons. These two conditions often conflict with each other, making  materials search difficult. 
Another proposal is low-dimensionality. 
\cite{Hicks93a} Low-dimensional 
systems have a peaked structure in the density of states, which is good for
large $S$. 
Despite these proposals, good thermoelectrics have remained 
elusive and 
awaits qualitatively new approaches for improvement of $ZT$.

In this Rapid Communication we propose that the quantum spin Hall (QSH) materials show
enhanced thermoelectric figure of merit at low temperature.
The QSH systems are new state of matters for bulk insulators
\cite{Kane05a,Bernevig06a,Murakami06a}, realized in two-dimension(2D) and 
in three-dimension(3D).
The 2D QSH system has gapless edge states which are stable against nonmagnetic impurities \cite{Wu06,Xu06a}.
Hence we expect that in dirty systems, electron conduction through the
edge states remain good, while phonon conduction is suppressed, satisfying
the PGEC criterion.
In addition, the edge states are one-dimensional(1D), which fits the
``low-dimensional" criterion.
Another good reason for this expectation 
is that 
the QSH effect was observed in Bi$_{1-x}$Sb$_x$\cite{Hsieh},  Bi$_2$Se$_3$\cite{Xia}, and 
Bi$_2$Te$_3$\cite{Chen} which are good thermoelectric materials. 

In 2D QSH systems in ribbon geometry, both the bulk 
states and edge states contribute. Because the number of bulk states is proportional to the ribbon 
width, we set the ribbon width to be very narrow, thereby the edge states
can have comparable or even larger contribution,
compared with the bulk. 
We then find that the
bulk and edge contributions compete each other. 
We also find that there occurs a bulk-to-edge crossover when the temperature is lowered. 
Because the edge states 
undergo inelastic scattering and lose their 
coherence, inelastic scattering 
length $\ell_{\mathrm{inel}}$ gives an effective system size for 
quantum transport by edge states.
As the temperature is lowered, $\ell_{\mathrm{inel}}$ become longer,  
and the edge states become dominant in thermoelectric transport.
We note that the edge transport cannot be dominant over the bulk 
transport at room temperature because 
$\ell_{\mathrm{inel}}$ might become very short.

The electric current $j$ and thermal current $w$ are coupled, and are induced 
by the thermal gradient or the electric field.
In a linear response, they are described as 
\begin{equation}
\left(\begin{array}{c}  
j/q \\ w \end{array}\right)=
\left(\begin{array}{cc} L_{0} & L_{1} \\ L_{1} & L_{2} \end{array}\right)
\left(\begin{array}{c}
 -\frac{\mathrm{d}\mu}{\mathrm{d}x} \\ -\frac{1}{T}\frac{\mathrm{d}T}{\mathrm{d}x} \end{array}\right),
\end{equation}
where $q$ is the electron charge $-e$, and $\mu$ is the chemical potential.
Thermal and electric properties are given by
\begin{eqnarray*}
&&\sigma=e^2L_0,\ \  \ S = -\frac{1}{eT}\frac{L_1}{L_0},\ \ \ 
\kappa_e = \frac{1}{T}\frac{L_0 L_2 - L_1^2}{L_0},\\
&&ZT = \frac{L_1^2}{L_0 L_2 - L_1^2 + \kappa_L T L_0}
\end{eqnarray*}
where $\kappa_e$ is the electron thermal conductivitiy, and $\kappa_L$
is phonon thermal conductivity.

\begin{figure}[htbp]
 \begin{center}
  \includegraphics[width=75mm]{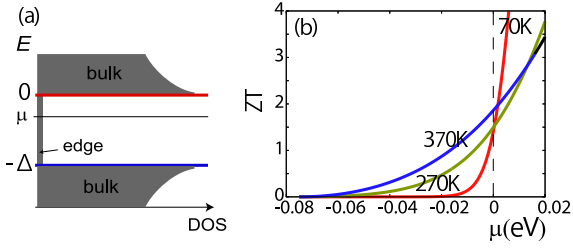}
 \end{center}
 \caption{(Color online) (a) Schematic bands for the bulk and edge states used in the calculation. (b) Thermoelectric figure of merit $ZT$ as a function 
of chemical potential $\mu$, by considering only the edge states in the 2D QSH system. Transport by bulk carriers and phonons is ignored.}
 \label{fig:edge}
\end{figure}
We first consider the edge transport only and neglect the bulk part. 
This corresponds to a case with very strong disorder, where 
the bulk states are assumed to be insulating, and 
the phonon heat transport is 
negligible. To describe the coherent transport of the edge states, we use the Landauer formula.
The density of states are schematically shown in
 Fig.~\ref{fig:edge}. The edge states are assumed to be perfectly conducting over the whole sample, and the transmission coefficient $T(E)$ is
unity when the electron energy is within the bulk gap 
($-\Delta < E < 0$). 
Here we measure 
the energy  from the 
bottom of the conduction band, and $\Delta$ is the energy gap. 
To clarify 
an interplay between the bulk and the edge states, we focus on the bottom of the bulk conduction band and neglect the valence band. 
We restrict the chemical potential to be $-\Delta/2\ll \mu$.
$L_\nu$ is given by
\begin{align}
L^{{\rm e}}_{\nu} =\frac{\ell}{sh}\int \mathrm{d}E T(E)(E-\mu)^{\nu} \left(-\frac{\partial f}{\partial E} \right),
\end{align}
where the suffix e means the edge transport, $h$ is the 
Planck constant. 
$\ell$ and $s$
are the length of the sample and the cross section of the sample.
This is rewritten as 
\begin{eqnarray}
L^{{\rm e}}_\nu =\frac{2\ell}{sh}(k_B T)^\nu \int^{-\bar{\mu}}_{-\bar{\Delta}-\bar{\mu}} x^{\nu} \frac{\mathrm{e}^x}{(\mathrm{e}^x+1)^2}  \mathrm{d}x
\label{eq:edge}
\end{eqnarray}
where  $\bar{\mu}= \frac{\mu}{k_BT}$, and $\bar{\Delta}= \frac{\Delta}{k_BT}$. We calculate $ZT$, by employing the gap size of Bi$_2$Te$_3$ ($\Delta=$0.15 eV).The result (Fig.~\ref{fig:edge}) shows that $ZT$ becomes larger and 
well exceeds unity, when the chemical 
potential is in the bulk band. It is because the edge states carry large energy.

In reality,  
when the chemical potential is in the bulk band, 
the bulk transport dominates,
and reduces $ZT$ from the otherwise large value. 
We treat the bulk and the edge transport 
independently, which is valid within the inelastic scattering length.
We calculate the bulk transport by 
the Boltzmann equation as
\begin{align}
L^{{\rm b}}_{\nu} =\int \mathrm{d}E\ (E-\mu)^{\nu} 
\left(-\frac{\partial f}{\partial E} \right) D(E)\tau,
\end{align}
where the suffix b means the bulk transport, and $D(E)$ is the density of states. $\tau$ is the relaxation time which is assumed to be constant.
The bulk band is assumed to be parabolic with an effective mass $m$. 
For simplicity, we include only the first subband
due to the confinement within the ribbon, by assuming that
 the gap between the first subband and second subband is large.  
The transport coefficients are then given by $L_\nu=L^{e}_\nu+L^{b}_\nu$ with
\begin{eqnarray}
L^{{\rm b}}_\nu =\frac{4\sqrt{2mk_B T}\mu^{*}c(k_B T)^\nu}{esh} 
\int^{\infty}_{-\bar{\mu}}\frac{ 
\sqrt{x+\bar{\mu}}x^\nu \mathrm{e}^{x}}{(\mathrm{e}^{x }+1)^2}  \mathrm{d}x
\label{eq:bulk}
\end{eqnarray}
where $\mu^{*}$ is the mobility, and 
the coefficient $c$ is the number of the carrier pockets. 

\begin{figure}[htbp]
 \begin{center}
  \includegraphics[width=85mm]{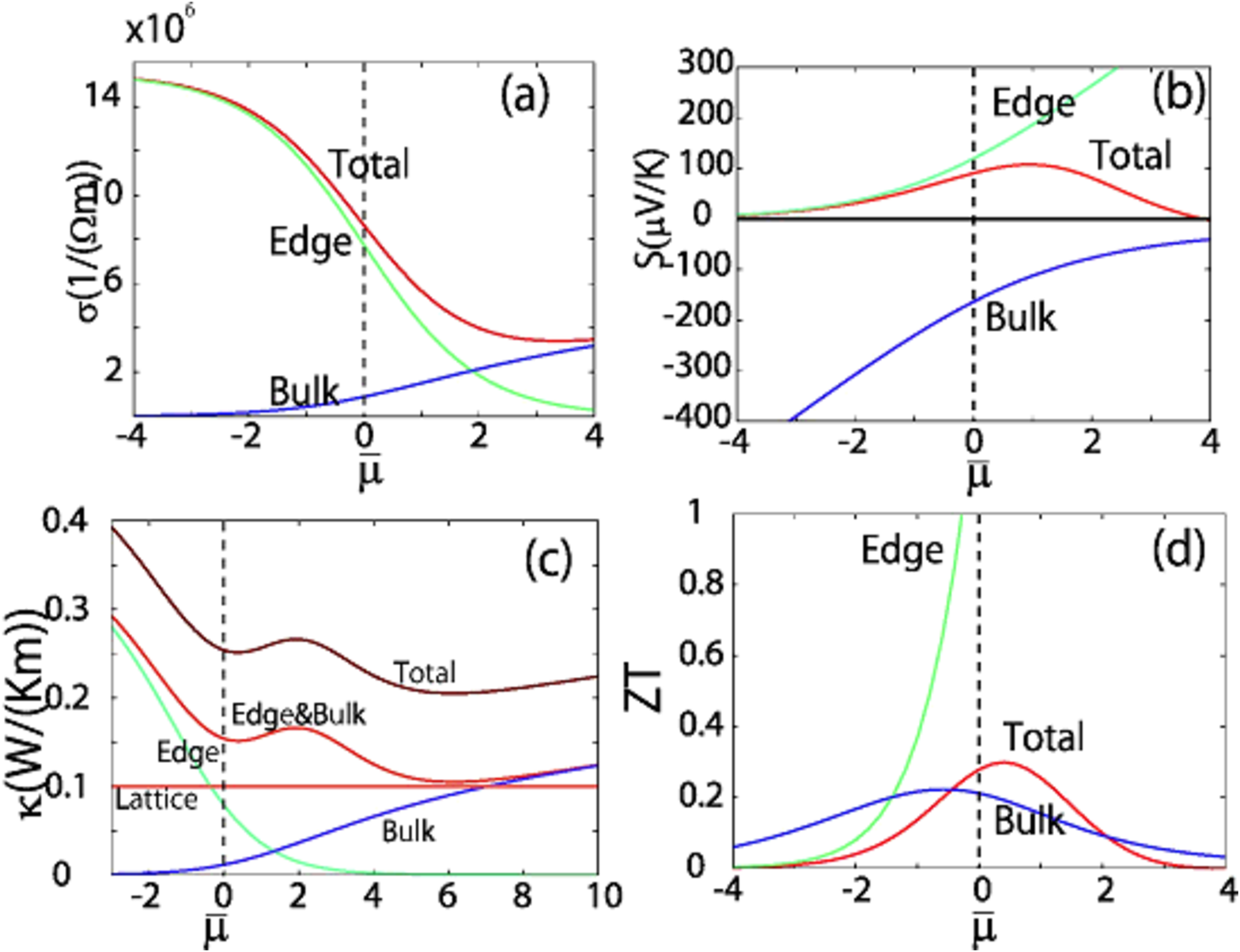}
 \end{center}
 \caption{(Color online) A calculation example of 
(a) Conductivity, (b) Seebeck coefficient,
(c) thermal conductivity, and (d) $ZT$ as a function of
the chemical potential.}
 \label{fig:properties}
\end{figure}

We calculate these transport properties at $T=$1.8K. We again employ the parameters for Bi$_2$Te$_3$ as follows. The parameters for bulk transport is  taken from those for bulk Bi$_2$Te$_3$. The electron effective mass is 0.02$m_e$ where $m_e$ is the electron mass, $c$ is 6. $\mu^{*}$ is measured at temperatures higher than 80K, and we estimated $\mu^{*}$ to be 2000cm$^2$V$^{-1}$s$^{-1}$ at $T=$1.8K by assuming that $\mu^{*}$ saturates at lower temperature due to disorder.
The effective system size $\ell$ is the inelastic scattering length $\ell_{\mathrm{inel}}$, and 
we assume $\ell\sim 1 \mu $m, which is a lower bound of $\ell_{\mathrm{inel}}$ in HgTe quantum well at 1.8K \cite{Roth}. $s$ is 10nm $\times$ 0.5 nm. 
$\kappa_L$ is 0.1 Wm$^{-1}$K$^{-1}$, which is expected from extrapolation from experimental data\cite{Macdonald} and theoretical estimate
\cite{Balandin}. These parameters might have some error bars, because of the lack of the experimental data for Bi$_2$Te$_3$ thin film. The results are shown in Fig.~\ref{fig:properties}.
For these parameters 
the energy difference between the first and
the second subbands is about 0.14 eV,
and
the chemical potential $\mu$ is assumed be less than this energy.
Many thermoelectric materials such as Bi$_2$Te$_3$ are narrow-gap semiconductors, and the effective mass is much smaller than the electron mass. Hence 
the subband structure is prominent, and the above assumption is 
satisfied without difficulty.

From Figs.~\ref{fig:properties}, $ZT$ has a maximum
when the chemical potential $\mu$ is near the band edge. This results from a
competition between the bulk and the edge states 
as follows. 
The Seebeck coefficient from the bulk states is larger when $\mu$ is in the bulk gap, whereas that 
from the edge states 
is larger when $\mu$ is in the bulk band. 
Their effects tend to cancel each other, because their charges have
opposite signs.
Therefore, maximum of $ZT$ occurs when $\mu$ is around 
the band edge.

\begin{figure}[htbp]
 \begin{center}
  \includegraphics[width=80mm]{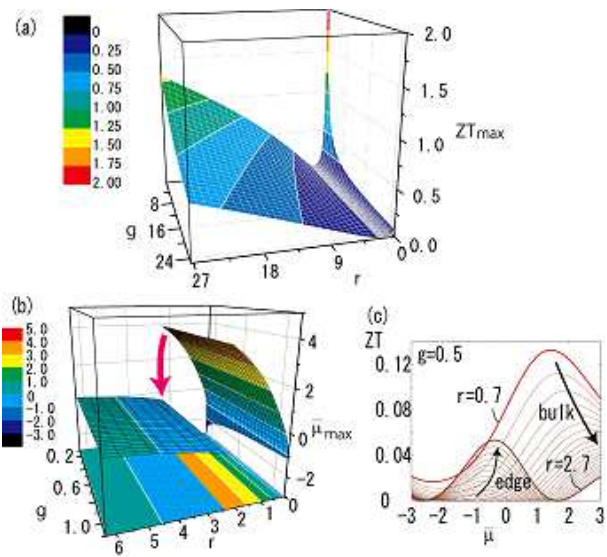}
 \end{center}
 \caption{(Color online) (a) $ZT_{\rm max}$ and (b) 
$\bar{\mu}_{\rm max}$ as a function of $r$ and $g$.
(c) $ZT$ as a function of $\bar{\mu}$ for various values of $r$ at $g=0.5$.
Bulk-to-edge crossover is seen by increasing $r$.}
 \label{fig:ZT3D}
\end{figure}

For optimization of the thermoelectric figure 
of merit in QSH systems, we define the following dimensionless 
parameters from the prefactors in Eqs.~(\ref{eq:edge}) and (\ref{eq:bulk});
\begin{align}
&r=\left[\frac{2\ell}{sh}\right]/\left[\frac{4\sqrt{2mk_B T}\mu^{*}c}{esh}\right]
=
\frac{e\ell}{2\sqrt{2mk_B T}\mu^{*}c},\ \\
&g=\left[\frac{\kappa_L}{k_B^2 T}\right]/\left[\frac{4\sqrt{2mk_B T}\mu^{*}c
}{esh}\right]
=\frac{\kappa_L esh}{4\sqrt{2mk_B^{5}T^{3}}\mu^{*}c}.
\end{align}
The parameter $r$ represents the ratio between the edge and the bulk transport
and $g$ represents the ratio between the phonon heat transport and the bulk transport. These ratios characterize thermoelectric transport 
of 2D QSH systems. 
For each $r$ and $g$ we maximize $ZT$ as a function of $\bar{\mu}$. In 
Fig.~\ref{fig:ZT3D}, 
we show the maximum $ZT_{\textrm{max}}$ and the value of $\bar{\mu}=
\bar{\mu}_{\textrm{max}}$ 
giving the maximum. To focus on an interplay between bulk and edge transport, we restrict $\mu$ to be near the conduction band edge, and ignore the valence band, by putting $\bar{\Delta}\to \infty$. 
From Fig.~\ref{fig:ZT3D}(a), 
as a function of $r$, $ZT_{\textrm{max}}$ becomes minimum 
at $r\sim 2.6$, because of a
competition between the edge- and bulk-state transport.
This interplay is prominent in the plot of 
$\bar{\mu}_{\textrm{max}}$ in Fig.~\ref{fig:ZT3D}(b). The plot has
a jump at around $r\sim 2.6$. 
As seen in Fig.~\ref{fig:ZT3D}(c), at about $r\sim 2.6$, the plot of $ZT$ as a function of $\bar{\mu}$ has two peaks, one from the bulk and the other 
from the edge. As $r$ passes through 2.6 from below, the peak from the edge dominates the peak from the bulk, and bulk-to-edge crossover occurs.

In Fig.~\ref{fig:ZT3D}(a) we can see
that at $r=0$ (no edge transport), the resulting $ZT$ is 
sensitive to $g$, and it is important to reduce $g$ by suppressing 
the phonon heat transport. 
However, disorder also suppresses electronic transport, and $ZT$ is
not enhanced so much.
On the other hand, as $r$ becomes larger, the result 
becomes insensitive to $g$. Disorder will enhance $r$, because 
the bulk mobility becomes smaller. The $ZT$ will then be enhanced.
Generally, at low temperatures both $r$ and $g$ tend to increase as $T$ decreases,
as we explain in the following. 
As $T$ is lowered, the mobility $\mu^{*}$ increases and eventually saturates.
$\kappa_L$ is given by $\kappa_L=\frac{1}{3}Cv_{L}l_{L}$, where $C$ is the
phonon specific heat, $v_L$ is the phonon velocity, and $l_{L}$ is
the phonon mean free path. As the temperature decreases, $l_{L}$ becomes
larger and saturates, while $C$ decreases; hence, $\kappa_L$ first increases and  then decreases at lower temperatures. From these behaviors, 
$r$ and $g$ tend to increase at low temperatures, possibly below around
10K. An estimation using the above-mentioned parameters for Bi$_2$Te$_3$ nanoribbon gives $r=9.4$ and $g=8.2$ at $T=1.8$K, which is located in the edge-dominated regime. 
We can estimate the crossover temperature for Bi$_2$Te$_3$ narrow ribbon 
taking into account the temperature dependence $\mu^{*}$ and $\kappa_L$ in the similar way as in Fig.~\ref{fig:properties}, and assuming that $\ell_{\mathrm{inel}}$, decreases as $T^{-1.5}$ as has been observed in quantum Hall systems \cite{Machida}. The crossover temperature is estimated to be around 5K-10K.

To realize the edge-dominated transport, 
the ribbon width $w$ should be much longer than the penetration depth 
$\lambda$ of the edge states, thereby we can ignore  hybridization  of the gapless edge states at the opposite edges. This hybridization induces a
gap $\delta\sim te^{-w/\lambda}$ to the edge states\cite{Zhou}, where $t$ is the bandwidth (several eV). The penetration depth $\lambda$ depends on the systems, and in some systems such as Bi ultrathin film, it is estimated to be on the order of the lattice constant \cite{Wada}. As we set $w=10{\rm nm}$ which is several decades of the lattice constant, the hybridization gap $\delta$ is estimated to be on the order of mK. Thus in our temperature range above 1K, this gap can be safely ignored. 
When we make the ribbon width to be much narrower, comparable to the penetration depth $\lambda$, the edge states at opposite edges hybridize and opens a sizable gap \cite{Zhou}, killing the perfectly conducting edge channels.

In 2D QSH systems, elastic backscattering of edge states due to nonmagnetic 
impurities is prohibited \cite{Wu06,Xu06a}. 
Inelastic scattering is a key factor to
characterize transport properties of the system. 
The electrons in 
edge states keep their coherence
within the inelastic scattering 
length $\ell_{{\rm inel}}$, which plays the role 
of the effective system size.
We first estimate the electron-phonon (el-ph) inelastic scattering length $\ell_{{\rm inel}}$,
 following the calculation on the quantum Hall (QH)
 system \cite{Zhao}. 
Here we assume the edge-state dispersion 
to be linear  with velocity $v_c$.
We put the bulk wavefunctions to be proportional 
to
$\mathrm{sin}(\pi y/w)$. 
By considering scattering by 2D longitudinal acoustic 
phonons, the relaxation time $\tau$ is given by 
$\tau^{-1} =(\tau^{ee})^{-1}+(\tau^{eb})^{-1}$,
where
$\tau^{ee}$, $\tau^{eb}$ are relaxation times by the edge-edge, the edge-bulk el-ph scattering. 
Following Ref.\cite{Zhao} we obtain
\begin{equation}
\frac{1}{\tau^{ee}}
\sim \frac{\pi V^2_{ep}T^2}{16\rho c^3_L v_c},\ \ 
\frac{1}{\tau^{eb}}
\sim \frac{\pi^3 V^2_{ep}T^2}{\rho c^3_Lv_c}\left(\frac{\lambda}{W}\right)^3,
\end{equation}
where $V_{ep}$ is a screened el-ph scattering
potential.
If we take $V_{ep} = 10^{-19}$J, $T = 1$K, $\rho = 10^{-6}$kg/m$^2$, $c_L = 10^{3}$m/s, $\lambda = 10^{-10}$m, $W=10^{-9}$m as an example, we get $\tau^{ee}\sim 10^{-8}$s, $\tau^{eb}\sim 10^{-6}$s. If $v_c\sim 10^6$m/s,
 $\ell_{\mathrm{inel}}=v_c \tau\sim 10^{-2}$m. Experimental $\ell_{\mathrm{inel}}$
is much shorter, implying that 
el-ph scattering is not crucial among
various inelastic scattering in the QSH system around 1K.

In addition to the el-ph interaction, 
the electron-electron (e-e) interaction also induces decoherence of edge 
states. There are two types of e-e interaction: edge-edge e-e 
and edge-bulk e-e interactions.
The edge-edge e-e interaction is renormalized into the edge state action,
and form the Luttinger liquid. Therefore this edge-edge e-e interaction
does not cause dephasing if the system is clean enough 
and the edge channels remain perfectly conducting well above 
Kondo temperature 
\cite{Maciejko}. 
In disordered systems, it gives rise to a finite
inelastic scattering time, while its estimate will be difficult.
In addition, 
the edge-bulk e-e interaction also appears at finite temperature, and 
it depends crucially on the details of the
system. 
Calculation of e-e interaction in the QSH systems is interesting
but is beyond the scope of the present Rapid Communication.

The inelastic scattering length $\ell_{\mathrm{inel}}$ is accessible 
experimentally. 
In the HgTe quantum well, nonlocal edge-state transport is observed
\cite{Roth} in 1$\mu$m sample at 1.8K. It indicates 
that $\ell_{\mathrm{inel}}$ is longer than the sample size, $\ell_{\mathrm{inel}}\ge 1\mu$m at $T = 1.8$K. It is limited 
by the potential inhomogeneity due to gating. 
On the other hand, the inelastic scattering length is measured in a QH system to be about 1$\mu$m at 1K \cite{Machida}, and is decreasing function of temperature. 
Based on these data we have used $\ell_{\mathrm{inel}}= 1\mu$m at $T=$ 1.8K in 
obtaining Fig.~\ref{fig:properties}, 
If the inelastic scattering length can be made longer, it will increase $r$
and enhance $ZT$ by edge-dominated thermoelectric transport. 

We address implications of our theory for 
3D QSH systems (topological insulators). 
Because the surface states on 3D QSH systems
are not perfectly conducting, the effect of 
surface states in 3D QSH systems on thermoelectriciy 
will be less prominent than that of edge states in 
2D QSH systems studied in this Rapid Communication.
Nevertheless, there can be one promising possibility also in the 3D QSH 
systems. 
In 3D QSH systems, 
protected 1D states \cite{Ran} of the crystal 
exist on line dislocations, depending on the bulk topological numbers. 
These 1D states are perfectly 
conducting. Recently, a prominent
 magnetofingerprint was observed in a topological insulator 
Bi$_2$Se$_3$, and it is suggested that the phase coherence is retained 
over 2mm at around 1K \cite{Checkelsky}. It is also suggested \cite{Checkelsky}, that
the transport involved in this magnetofingerprint is carried by 
these 1D states on dislocations.
If this scenario is true, they can be dominant in low temperatures, 
as we have shown 
in this Rapid Communication. The estimated phase coherence length 
$\ell_{\mathrm{inel}}\sim$2mm is three orders of 
magnitude larger than that we used in our calculation, 
and it is favorable
for thermoelectric transport.  

Recently, an anomalous enhancement of the Seebeck coefficient at 7K is reported in p-Bi$_{2}$Se$_{3}$ \cite{Hor}.
Though our 2D model cannot describe three-dimensional p-Bi$_{2}$Se$_{3}$, we may attribute this enhancement to 
either surface states or 1D states along line dislocations. 
In particular, the 1D states form perfectly conducting channels, and
will enhance the figure of merit.
We note that 
in our calculation the edge and bulk contributions to the Seebeck coefficient
has opposite signs, because the carrier charges have opposite signs
(i.e.\ holes and electrons), and therefore the Seebeck coefficient
changes sign at the bulk-to-edge crossover by changing $T$. 
On the other hand, the Seebeck coefficient on p-Bi$_{2}$Se$_{3}$
does not change sign by lowering temperature. Within our interpretation 
this implies that the bulk carriers and the 1D carriers have the
same signs for the charge in the experiment. 

To summarize, we study thermoelectric properties of two-dimensional 
quantum spin Hall systems. The edge states become dominant in thermoelectric 
transport at low temperature, which might be below 5K-10K for narrow ribbons.
This bulk-to-edge crossover temperature is higher for longer inelastic scattering length of edge states. 

We are grateful to T.~Machida, X.~-L.~Qi, and S.~-C.~Zhang for helpful discussions. 
This research is supported in part 
by Grant-in-Aids  
from MEXT.


\end{document}